\def\year{2019}\relax
\documentclass[letterpaper]{article} 
\usepackage{aaai19}  
\usepackage{times}  
\usepackage{helvet}  
\usepackage{courier}  
\usepackage{url}  
\usepackage{graphicx}  
\usepackage{multirow}
\usepackage{color}
\usepackage{amsmath}
\usepackage{amssymb}
\usepackage{subcaption}

\begin{document}
%
\title{PerformanceNet:
Score-to-Audio Music Generation with Multi-Band \\Convolutional Residual Network
}

\author{Bryan Wang and  Yi-Hsuan Yang\\
Research Center for Information Technology Innovation, Academia Sinica, Taipei, Taiwan\\
{\{bryanw, yang\}}@citi.sinica.edu.tw\\
}

\maketitle
\begin{abstract}
Music creation is typically composed of two parts: composing the musical score, and then performing the score with instruments to make sounds.
While recent work has made much progress in automatic music generation in the symbolic domain, few attempts have been made to build an AI model that can render realistic music audio from musical scores. Directly synthesizing audio with sound sample libraries often leads to mechanical and deadpan results, since musical scores do not contain performance-level information, such as subtle changes in timing and dynamics.
Moreover, while the task may sound like a text-to-speech synthesis problem, 
there are fundamental differences since music audio has rich polyphonic sounds.
To build such an AI performer, we propose in this paper a deep convolutional model that learns in an end-to-end manner the score-to-audio mapping between a symbolic representation of music called the pianorolls and an audio representation of music called the spectrograms. 
The model consists of two subnets: the \emph{ContourNet}, which uses a U-Net structure to learn the correspondence between pianorolls and spectrograms and to give an initial result; 
and the \emph{TextureNet}, which further uses a multi-band residual network to refine the result by adding the spectral texture of overtones and timbre.
We train the model to generate music clips of the violin, cello, and flute, with a dataset of moderate size.
We also present the result of a user study that shows our model achieves higher mean opinion score (MOS) in naturalness and emotional expressivity than a WaveNet-based model and two off-the-shelf synthesizers. We open our source code at \url{https://github.com/bwang514/PerformanceNet}
\end{abstract}

\section{Introduction}

Music is an universal language used to express the emotion and feeling beyond words. We can easily enumerate a number of commonalities between music and language. 
For example, the musical score and text are both the symbolic transcript of their analog counterparts---music and speech, which manifest themselves by sounds organized in time. 

However, a fundamental difference that distinguishes music from every other languages is that the essence of music exists mainly in its audio performance \cite{widmer16tist}. Humans can comprehend the meaning of language by reading text.
But, we cannot fully understand the meaning of a music piece by merely reading its musical score. 
Music performance is necessary to give music meanings and feelings. 
In addition, musical score only specifies ‘what’ notes to be played instead of elaborating ‘how’ to play them. Musicians can leverage this freedom to interpret the score and add expressiveness in their own ways, to render unique pieces of sound and to ``bring the music to life'' \cite{raphael09ijcai}.


The space left for human interpretation gives music the aesthetic idiosyncrasy, but it also makes automated music synthesis, 
the process of converting a musical score to audio by machine, a challenging task.
While modern music production industry has developed sophisticated digital systems for music synthesis,
using the default synthesis parameters would still often lead to deadpan results. 
One has to fine-tune the synthesis parameters for each individual note, sometimes with laborious trial-and-error experiments, to produce a realistic music clip \cite{riionheimo03jasp}. 
These parameters include dynamics attributes such as velocities and timing, as well as timbre attributes such as effects.  
This requires a great amount of time and domain expertise. 

One approach to automate the fine-tuning process is to train a machine learning model to predict the expressive parameters for the synthesizers \cite{macret14gecco}. 
However, such an auto-parameterization approach may work better for instruments that use relatively simple mechanisms to produce sounds, e.g., striking instruments such as the piano and drums. Instruments that produce sounds with more complicated methods, such as the bowing string instruments and the aerophones, are harder to be comprehensively parameterized by pre-defined attributes.  

To address this problem, we propose to directly model the end-to-end mapping between musical scores and their audio realizations with a deep neural network, 
forcing the model to learn its own way the \emph{audio attributes} that are important and the mechanism to fine-tune them. 
We believe that this is a core task in building an ``AI performer''---to learn how a human performer renders a musical score into audio. Other tasks that are important include modeling the style and mental processes of the human performers (e.g., the emotion they try to express), and using these \emph{personal attributes} to condition the audio generation process. 
We attempt to learn only the score-to-audio mapping here, leaving the ability to condition the generation a future work.

Text-to-speech (TTS) generation has been widely studied in the literature, with WaveNet and its variants being the state-of-the-art \cite{wavenet,shen2018tacotron2}.
However, music is different from speech in that music is polyphonic (while speech is monophonic), and that the musical score can be viewed as a time-frequency representation (while the text transcript for speech cannot). 
Since music is polyphonic, we cannot directly apply a TTS model to score-to-audio generation.
But, since the score contains time-frequency information, we can exploit this property to 
design a new model for our task.

\begin{figure}
\begin{center}
\includegraphics[width=.475\columnwidth]{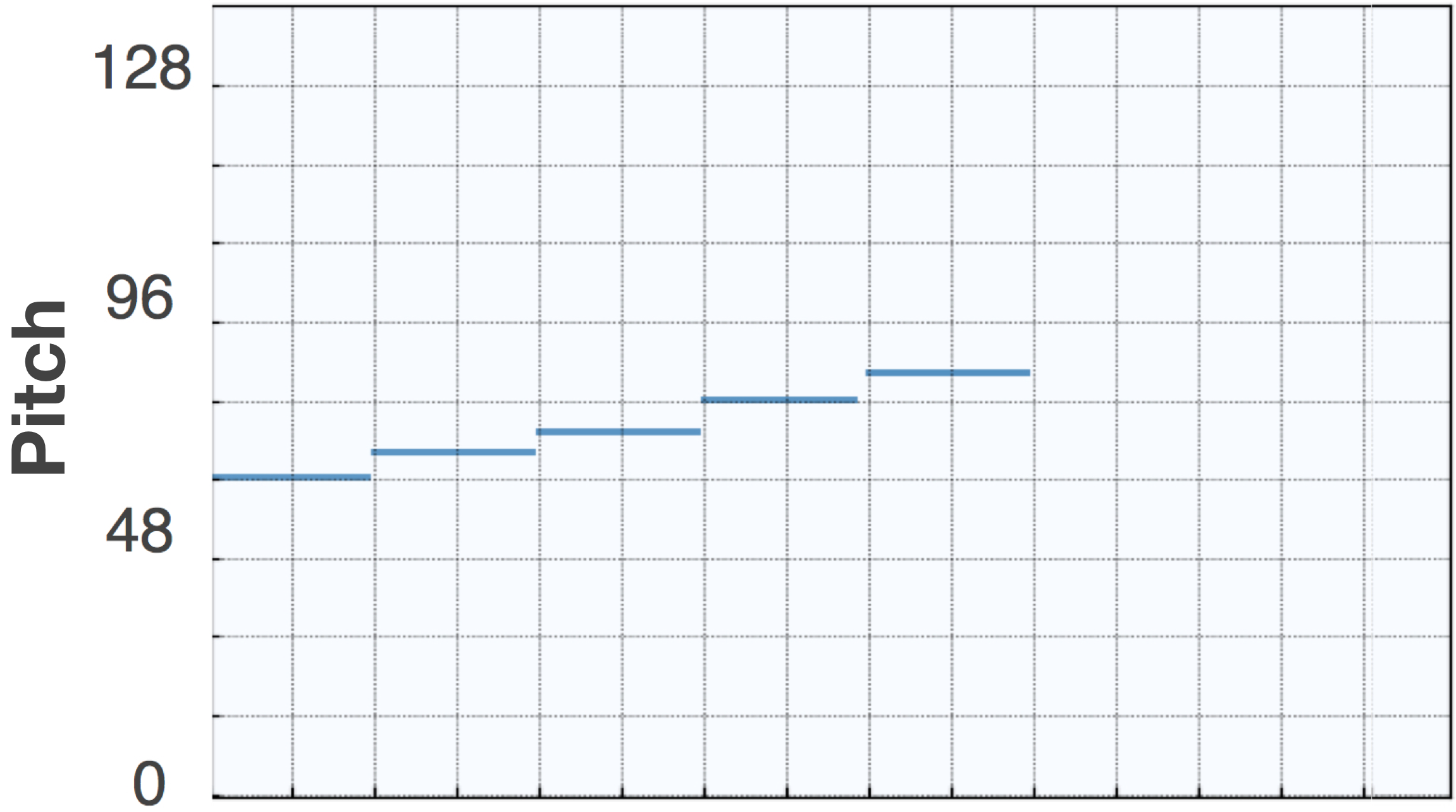}
\includegraphics[width=.51\columnwidth]{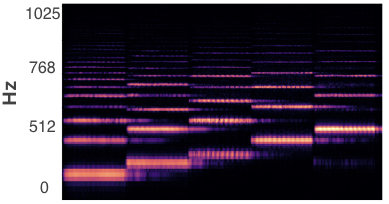}
\caption{The spectrogram of a music clip (right) and its corresponding pianoroll (left).}
\label{fig:pianoroll}
\end{center}
\end{figure}

A musical score can be viewed as a time-frequency representation when it is represented by the \emph{pianorolls} \cite{pypianoroll}.
As exemplified in Figure~\ref{fig:pianoroll}, a pianoroll is a binary, scoresheet-like matrix representing the presence of notes over different time steps for a single instrument. When there are multiple instruments, we can extend it to a tensor, leading to a multitrack pianoroll. A multitrack pianoroll would be a set of pianorolls, one for a different track (instrument).
On the other hand, we can represent an audio signal by the spectrogram, a real-valued matrix representing the spectrum of frequencies of sounds over time. Figure~\ref{fig:pianoroll} shows the nice correspondence between a (single-track) pianoroll and its spectrogram---the presence of a note in the pianoroll would incur a harmonic series (i.e., the fundamental frequency and its overtones) in the spectrogram. 
Given paired data of scores and audio, we can exploit this correspondence to build a deep convolutional network that convert a matrix (i.e., the pianoroll) into another matrix (i.e., the spectrogram), to realize score-to-audio generation.
This is the core idea of the proposed method.

A straightforward implementation of the above idea, however, cannot work well in practice, because the pianorolls are much smaller than the spectrograms in size. 
The major technical contribution of the paper is a new model that combines convolutional encoders/decoders and multi-band residual blocks to address this issue. We will talk about the details of the proposed model in the Methodology Section.

To our best knowledge, this work represents the first attempt to achieve score-to-audio generation using fully convolutional neural networks.
Compared to WaveNet, the proposed model is less data-hungry and is much faster to train.
For evaluation, we train our model to generate the sounds of three different instruments 
and conduct a user study.
The ratings from 156 participants show that our model performs better than a WaveNet-based model \cite{manzelli18ismir} and two off-the-shelf synthesizers in the the mean opinion score  of naturalness and emotional expressivity.


\begin{figure}[t]
\begin{subfigure}[c]{\linewidth}\centering
\includegraphics[width=.7\columnwidth]{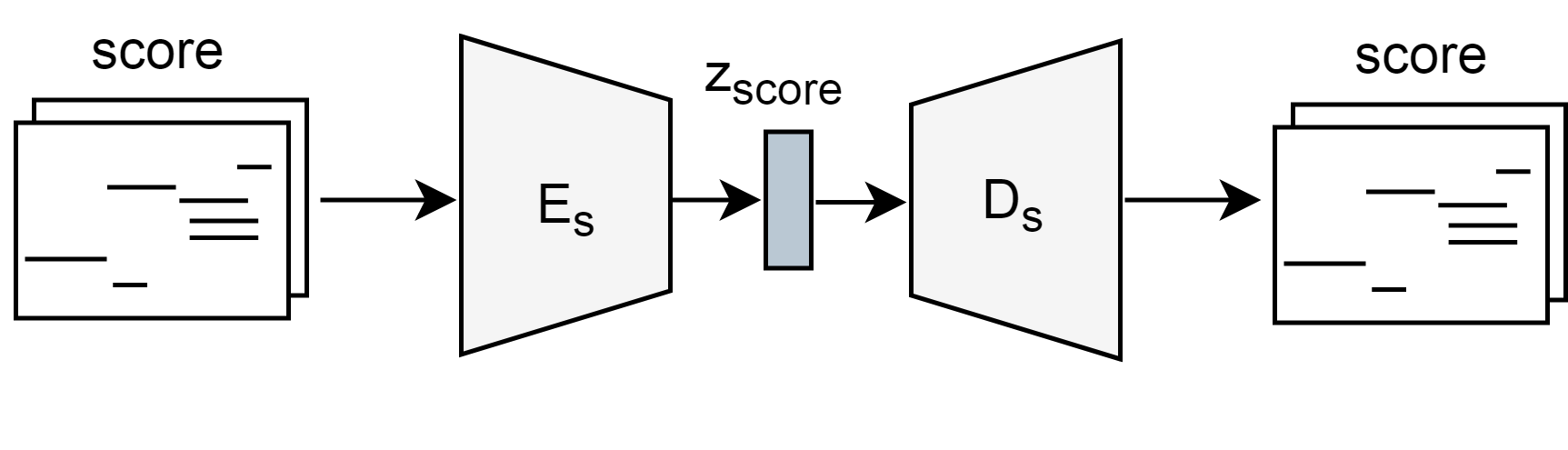}
\caption{Score-to-score model}\label{fig:s2s}
\end{subfigure}
\begin{subfigure}[c]{\linewidth}\centering
\includegraphics[width=.7\columnwidth]{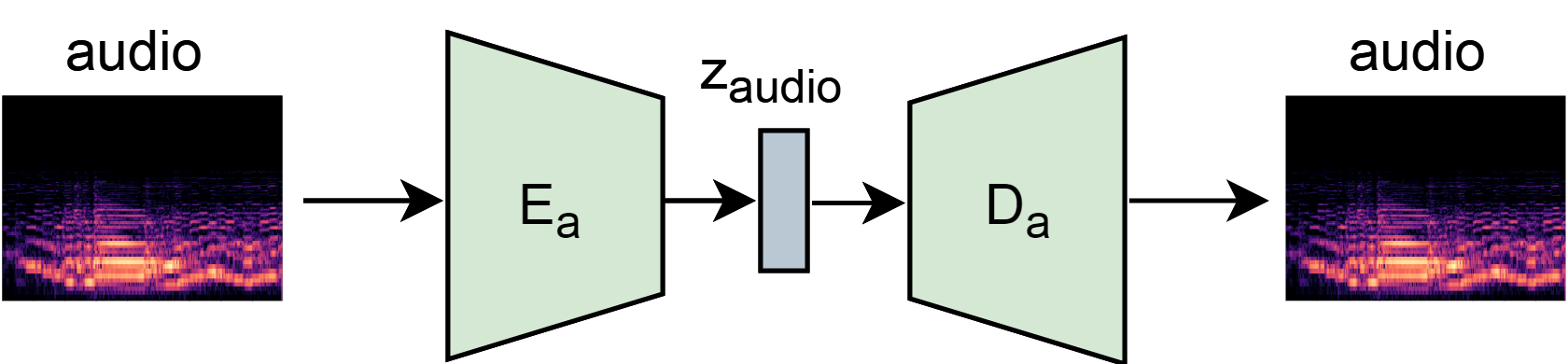}
\caption{Audio-to-audio model}\label{fig:a2a}
\end{subfigure}
\begin{subfigure}[c]{\linewidth}\centering
\includegraphics[width=.7\columnwidth]{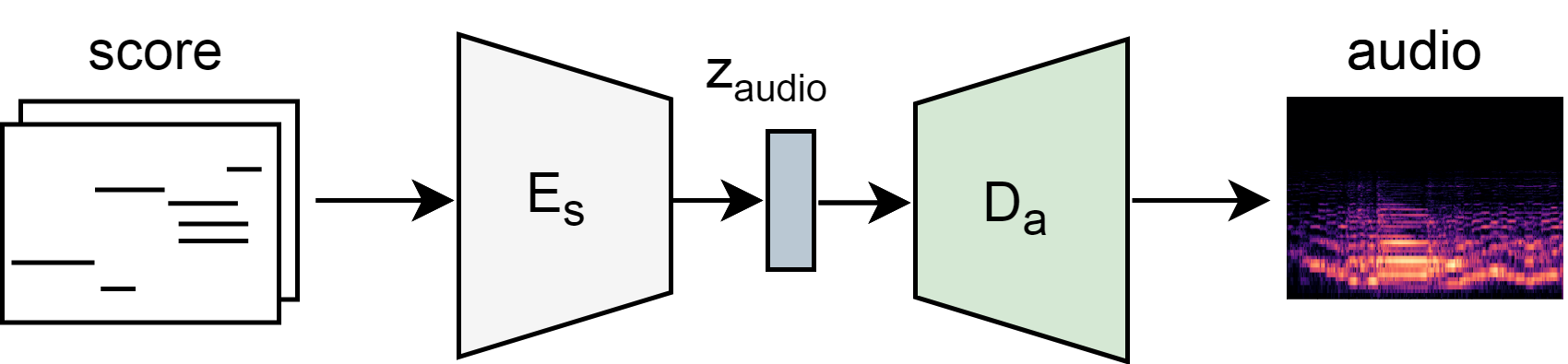}
\caption{Score-to-audio model}\label{fig:s2a}
\end{subfigure}
\begin{subfigure}[c]{\linewidth}\centering
\includegraphics[width=.95\columnwidth]{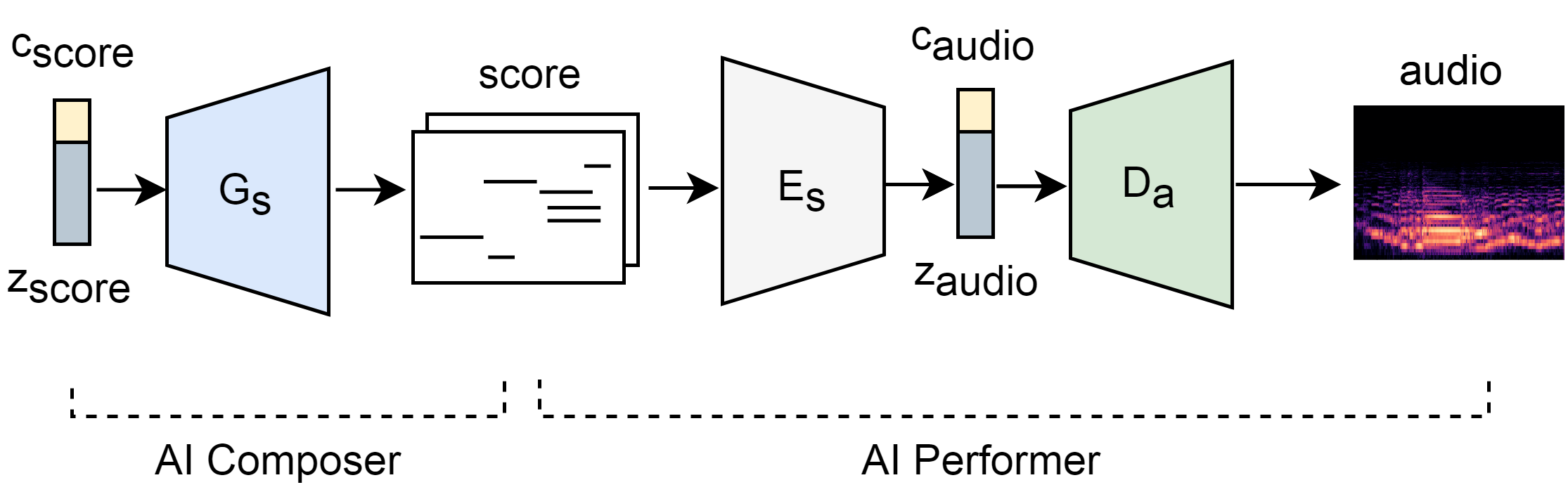}
\caption{AI Composer $+$ AI Performer }\label{fig:a2a}
\end{subfigure}
\caption{Illustration of the encoder/decoder network for different tasks. (Notation:  $E$, $D$, $\mathbf{z}$, and $\mathbf{c}$ denote the encoder, decoder, latent code, and condition code, respectively.)}
\label{fig:related}
\end{figure}

\section{Background}

\subsection{Music Generation}
Algorithmic composition of music
has been studied for decades. 
It garners remarkably increasing attention in recent years along with the resurgence of AI research. Many deep learning models have been proposed to build an AI composer. 
Among existing works, MelodyRNN  \cite{melodyRNN16} and DeepBach  \cite{deepbach} are arguably the 
most well-known.
They claim that they can generate realistic melodies and Bach chorales, respectively. 
Follow-up research has been made to generate duets \cite{AI_duet}, lead sheets \cite{roberts18musicVAE}, 
multitrack pianorolls, \cite{dong2017musegan,simon18ismir}, 
or lead sheet arrangement \cite{liu2018lead}, 
to name a few.
Many of these models are based on either 
generative adversarial networks (GAN) \cite{gan}
or variational autoencoders (VAE) \cite{vae}.

Although exciting progress is being made, we note that the majority of recent work on music generation concerns with only symbolic domain data, not the audio performance.
There are attempts to generate melodies with performance attributes such as the PerformanceRNN \cite{performancernn}, but it is still in the symbolic domain and it concerns with only the piano.
Broadly speaking, such \emph{score-to-score} models involve learning the latent representation (or latent code) of symbolic scores, as illustrated in Figure \ref{fig:related}(a).

A few attempts have been made to generate musical audio. A famous example is NSynth \cite{nsynth}, which uses VAE to learn a latent space of musical notes from different instruments. The model can create new sounds by interpolating 
the latent codes. There are some other models that are based on WaveNet \cite{dieleman18arxiv} or GAN \cite{donahue18arxiv,wu18icmlw}. Broadly speaking, we can illustrate such \emph{audio-to-audio} models with Figure \ref{fig:related}(b).

The topic addressed in this paper is expressive music audio generation (or synthesis) from musical scores, which involves a \emph{score-to-audio} mapping illustrated in Figure \ref{fig:related}(c). 
Expressive music synthesis has a long tradition of research \cite{widmer04jnmr}, 
but few attempts have been made to apply deep learning to score-to-audio generation for arbitrary instruments. 
For example, \cite{hawthorne2018enabling} deals with  the piano only.
One notable exception is the work presented by \cite{manzelli18ismir}, which uses WaveNet to render pianorolls. However, WaveNet is a general-purpose model for generating any sounds. We expect our model can perform better, since ours is designed for music. 

Figure \ref{fig:related}(d) shows that the score-to-score symbolic music generation model (dubbed the AI composer) can be combined with the score-to-audio generation model (dubbed the AI performer) to mimic the way human beings generate music.
This figure also shows that, besides the latent codes, we can add the so-called ``condition codes'' \cite{DCGAN} to condition the generation process. For the symbolic domain, such conditions can be the target musical genre. For the audio domain, such conditions can be some ``personal attributes'' of the AI performer, such as its personality, style, or the emotion it intends to express or convey  in the performance. 
We treat the realization of such a complete model as a future work.

\subsection{Deep Encoder/Decoder Networks}

We adopt a deep encoder/decoder network structure in this work for score-to-audio generation. We present some basics of such a network structure below.

The most fundamental and well-known encoder/decoder architecture is the autoencoder (AE) \cite{masci11icann}. 
In AE, the input and target output are the same. 
The goal is to compress the input  into a low-dimensional \emph{latent code}, and then uncompress the latent code to reconstruct the original input, as shown in Figures \ref{fig:related}(a) and (b). 
The data compression is achieved with a stack of downsampling layers known collectively as the \emph{encoder}, and the uncompression is made with a few upsampling layers, the \emph{decoder}.
The encoders and decoder can be implemented with either fully-connected, convolutional or recurrent layers.
When the input is a noisy version of the target output, the network is called a denoising AE. 
When we further regularize the distribution of the latent code to follow a Gaussian distribution (which facilitates sampling and accordingly data generation), the network is called a VAE \cite{vae}. 


A widely-used design for an encoder/decoder network is to add links between layers of the encoder and the decoder, by concatenating the output of an  encoding layer to that of a decoding layer.
While training the model, the gradients can be propagated directly through such \emph{skip connections}. This helps mitigate the gradient vanishing problem and helps train deeper networks. 
People also refer to an  encoder/decoder network that has skip connections as a U-net \cite{ronneberger2015u}.


\begin{figure*}[h]
\begin{center}
\includegraphics[width=.9\textwidth]{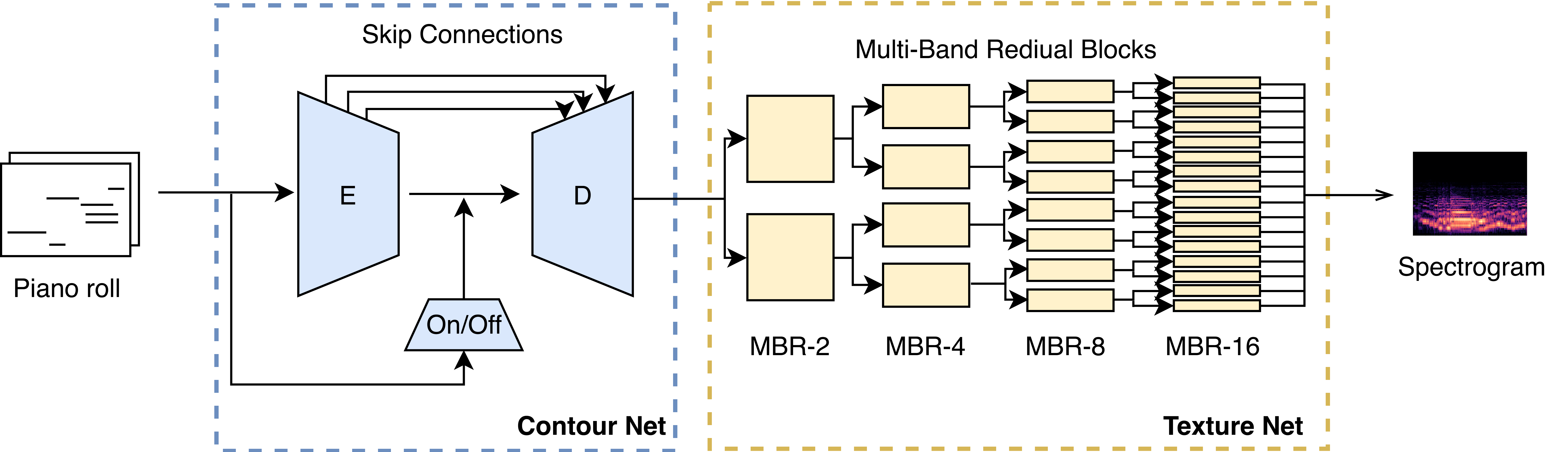}
\caption{System diagram of the proposed model architecture for score-to-audio generation. (Notation: On/Off and MBR-$k$ denote the onset/offset encoder and a multi-band residual block that divides a (log-scaled) spectrogram into $k$ bands, respectively.)}
\label{fig:sys}
\end{center}
\end{figure*}

\section{Methodology}

\subsection{Overview of the Proposed PerformanceNet Model}

A main challenge of score-to-audio generation using a convolutional network is that the pianorolls are smaller than the spectrograms in size. 
For example, in our implementation, the size of a (single-track) pianoroll is 128$\times$860, whereas the size of the corresponding spectrogram is 1,025$\times$860.
Moreover, a pianoroll is a binary-valued matrix with few number of 1's (i.e., it is fairly sparse), whereas a spectrogram is a dense, real-valued matrix.

To make an analogy to computer vision problems, learning the mapping between pianorolls and spectrograms can be thought of as an image-to-image translation task \cite{gatys16cvpr,liu17nips}, 
which aims to maps a sketch-like binary matrix to an image-like real-valued matrix. 
On the other hand, the challenges involved in converting a low-dimensional pianoroll to a high-dimensional spectrogram can also be found in image super-resolution problems \cite{dong14eccv,ledig17cvpr}. 
In other words, we need to address image-to-image translation and image super-resolution at the same time.

To deal with this challenge, we propose a new network architecture that consists of two subnets, one for image-to-image translation and the other for image super-resolution. 
The model architecture is illustrated in Figure \ref{fig:sys}.
\begin{itemize}
    \item The first subnet, dubbed the \emph{ContourNet}, deals with the image-to-image translation part and aims to generate an initial result. It uses a convolutional encoder/decoder structure to learn the correspondence between pianorolls and spectrograms shown in Figure \ref{fig:pianoroll}. 
    \item The second subnet, dubbed the \emph{TextureNet}, deals with image super-resolution and aims to enhance the quality of the initial result by adding the spectral texture of overtones and timbre. 
    \item After TextureNet, we use the classic Griffin-Lim algorithm \cite{griffin84} to estimate the phase of the spectorgram and then create the audio waveforms.
\end{itemize}


Specifically, being motivated by the success of the progressive generative adversarial network (PGGAN) in generating high resolution images \cite{karras18iclr}, we attempt to improve the resolution of the generated spectrogram layer-by-layer in the TextureNet with a few \emph{residual blocks} \cite{he16cvpr}.  
However, unlike PGGAN, we use a \emph{multi-band} design in the TextureNet and aim to improve only the frequency resolution (but not the temporal resolution) progressively.
This design is based on the assumption that the ContourNet has learned the temporal characteristics of the target and hence the TextureNet can focus on improving the spectral texture of overtones.
Such a multi-band design has been shown effective recently for blind audio source separation \cite{takahashi17waspaa}.

Moreover, to further encode the exact timing and duration of the musical notes, we add an additional encoder whose output is connected to the bottleneck layer of the ContourNet to provide the onset and offset information of the pianorolls. 

In what follows, we present the details of its two component subnets, the ContourNet and the TextureNet.

\subsection{Data Representation}

The input data of our model is a pianoroll for an instrument, and the target output is the spectrogram of the corresponding audio clip. 
A pianoroll is a binary matrix representing the presence of notes over time. 
It can be derived from a MIDI file \cite{pypianoroll}.
We consider $N=128$ notes for the pianorolls in this work.
On the other hand, the spectrogram of an audio clip is the magnitude part of its short-time Fourier Transform (STFT). Its size depends on the window size and hop size of STFT, and the length of the audio clip.  
In our implementation, the pianorolls and spectorgrams are time-aligned. When the size of a pianoroll is $N \times T$, the size of the corresponding spectrogram would be $F \times T$, where $F$ denotes the number of frequency bins in STFT. Typically, $N\ll F$.
Since a pianoroll may have variable length, we adopt a \emph{fully-convolutional} design \cite{oquab15localization} and use no pooling layer at all in our network. Yet, for the convenience of training the models with mini-batches, in the training stage 
we cut the waveforms in the training set to fixed length chunks of the same $T$.

\subsection{ContourNet}

\subsubsection{Convolutional U-net with an Asymmetric Structure}
As said, we use a convolutional encoder/decoder structure for the ContourNet. Specifically, as illustrated in Figure \ref{fig:sys}, we add skip connections between the encoder (`E') and the decoder (`D') to make it a U-net structure.
We find that such a U-net structure works well in learning the correspondence between a pianoroll and a spectrogram.
Specifically, it helps the encoder communicate with the decoder information regarding the pitch and duration of notes, which would otherwise be lost or be vague in the latent code, if we do not use the skip connections. 
Such detailed pitch information is needed to render coherent music audio. 

Moreover, since the dimensions of the pianorolls and spectrograms do not match, we adopt an \emph{asymmetric} design for the U-net so as to increase the frequency dimension gradually from $N$ to $F$. 
Specifically, we use a U-Net with a depth of 5 layers. 
In the encoder, we use 1D convolutional filters to compress the input pianoroll along the time axis 
and gradually increase the number of channels (i.e. the number of feature maps) by a factor of 2 per layer, from $N=128$ up to 4,096 to reach the bottleneck layer. 
In the decoder, we use 1D deconvolutional filters to gradually uncompress the latent code along the time axis, but decrease the number of channels from 4,096 to $F$. When the number of channels is equal to $F$ for a layer in the decoder,  we no longer decrease the number of channels for the subsequent layers, to ensure the output has the frequency dimension we desire.

\begin{figure*}[t]
\begin{center}
\includegraphics[width=.9\textwidth]{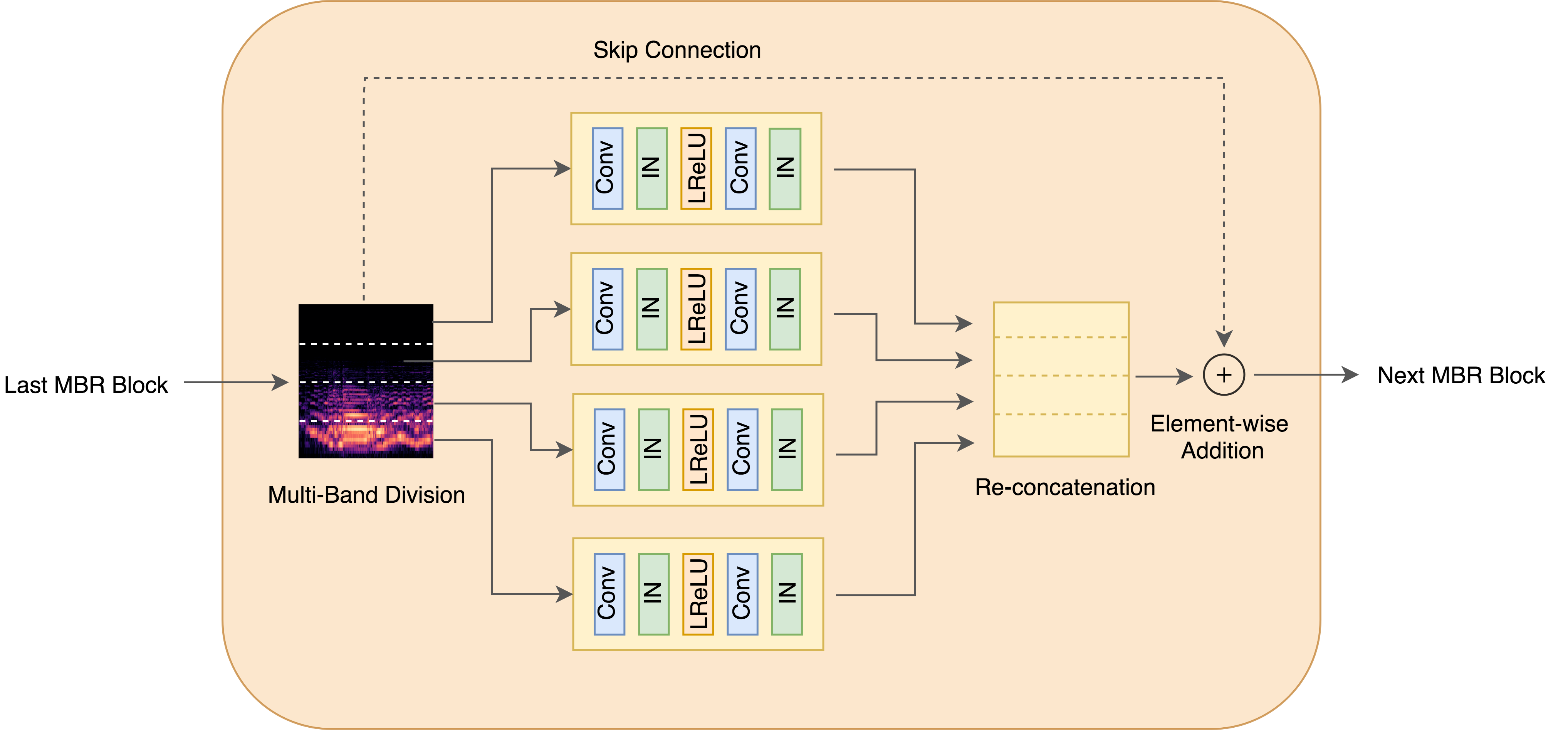}
\caption{Illustration of a multi-band residual (MBR) block in the proposed TextureNet. In each MBR block, the input spectrogram is split into a specific number of frequency bands. We then feed each band individually to identical sub-blocks consisting of the following five layers: 1D-convolution, instance norm \cite{ulyanov17cvpr}, leaky ReLU, 1D-convolution, and instance norm.
The output of all the sub-blocks are then concatenated along the frequency dimension and then summed up with the input of the MBR block. We then pass the output to the next MBR block. }
\label{fig:MBRBlock}
\end{center}
\end{figure*}

\subsubsection{Onset/offset Encoder}
Though the model presented above can already generate non-noisy, music-like audio, in our pilot studies on generating cello sounds, we find that it tends to perform \emph{legato} for a note sequence, even though we can see clear boundary or silent intervals between the notes from the pianoroll. 
In other words, this model always plays the notes smoothly and connected, suggesting that it has no sense of the offsets and does not know when to  stop the notes.

To address this issue, we improve the model by adding an encoder to incorporate the onset and offset information of each note from the pianoroll. 
This encoder is represented by the block marked `On/Off' in Figure \ref{fig:sys}.
It is implemented by two convolutional layers.
Its input is an `onset/offset roll' we compute from the pianoroll---we use $+1$ to represent the time of an onset, $-1$ to represent the offset, and $0$ otherwise. 
The onset/offset roll is even sparser than the pianoroll, but it makes onset/offset information explicit.
Moreover, we also use skip connections here and concatenate the output of these two layers to the beginning layers of the U-Net decoder, to inform the decoder the note boundaries during the early stage of spectrogram generation. 

Our pilot study shows that the information of the note boundaries greatly helps the ContourNet to learn how to play a note. For example, when playing the violin, the sound at the onset time of the note (i.e., when the bow contacts the string) is different from the sound for the rest of the time (i.e., during the time period the note is played). We can divide the sound of the note into three stages: it first starts with a thin timbre and low volume, gradually develops its velocity (i.e., energy) to the climax with obvious bowing sound, and then vanishes. The onset/offset encoder marks these different stages and makes it easier for the decoder to learn the subtle dynamics of music performance.

\subsection{TextureNet}
While the ContourNet alone can already  generate audio with temporal dependency and coherence, it fails to generate the details of the overtones, leading to unnatural sounds with low audio quality. 
We attribute this to the fact that we use the same convolutional filters for all the frequency bins in ContourNet. 
As a result, it cannot capture the various local spectral textures in different frequency bins. 
For instance, maybe the sounds from the high frequency bins should be sharper than those from the low frequency bins.

To address this issue, we propose a progressive multi-band residual learning method to better capture the spectral features of musical overtones. 
The basic idea is to divide the $F\times T$ spectrogram along the frequency axis into several smaller matrices $F' \times T$, each corresponding to a frequency band.
To learn the different characteristics in different bands, we use different convolutional filters for different bands, and finally concatenate the result from different bands to have the original size. 
Instead of doing this multi-band processing once, we do it progressively---we divide the spectrogram into fewer bands in the earlier layers of TextureNet, and into more bands in the latter layers of TextureNet, as Figure \ref{fig:sys} shows. In this way, we can gradually improve the result in different frequency resolution.

Specifically, similar to PGGAN \cite{karras18iclr}, we train different layers of TextureNet with  \emph{residual learning} \cite{he16cvpr}. 
As illustrated in Figure \ref{fig:MBRBlock}, this is implemented by adding a skip connection between the input and output of a \emph{multi-band residual block} (MBR), and use element-wise addition to combine the input with the result of the band-wise processing inside an MBR. With this skip connection, the model may choose to ignore the band-wise processing inside an MBR. In consequence, we can use several MBRs for progressive learning, each MBR divides the spectrogram into a certain number of bands.

We argue that the proposed design has the following two advantages. 
First, 
compared to separately generating the frequency bands, residual learning reduces the divergence among the generated bands, since the result of the
band-wise processing inside an MBR would be integrated with the original input of the MBR, which has stronger coherence among the whole spectrogram.
Second, by progressively dividing the spectrogram into smaller bands, the receptive field of each band becomes smaller. This helps the convolutional kernels in TextureNet to capture the spectral features from a wider range to a narrower range. The coarse-to-fine process is analogous to the way human painters paint--- oftentimes, painters first draw a coarse sketch, then gradually add the details and put on the colors layer-by-layer.

We note that the idea of multi-band processing has been shown effective by \cite{takahashi17waspaa} for blind audio source separation, whose goal is to recover the sounds of the audio sources that compose an audio mixture. 
However, they do not use a progressive-growing strategy.
On the other hand, in PGGAN \cite{karras18iclr}, they progressively enhance the quality of an image by improving the resolution along the two axes simultaneously.
We propose to do so along the frequency dimension only, so that ContourNet and Texture can have different focuses.

Given that processing different bands separately may lead to artifacts at the boundary of the bands, \cite{takahashi17waspaa} propose to split the frequency bands with overlaps and apply a hamming window to 
concatenate the overlapping areas. 
We do not find this improves our result much.


\section{Dataset}


We train our model with the MusicNet dataset \cite{thickstun2017learning}, which provides  time-aligned MIDIs and audio clips 
for more than 34 hours of chamber music. 
We convert the MIDIs to pianorolls by the \texttt{pypianoroll} package \cite{pypianoroll}.
The audio clips are sampled at 44.1 kHz.
To have fixed-length input for our training data, we cut the clips into 5-second chunks. 
We compute the log-scaled spectrogram with 2,048 window size and 256 hop size with the \texttt{librosa} package \cite{librosa}, leading to a 1,025$\times$860 spectrogram and 128$\times$860 pianoroll for each input/output data pair.
We remark that it is important to use a small hop size for the audio quality of the generated result. 

 
Due to the differences in timbre, we need to train a model for each instrument. 
From MusicNet, we can find solo clips for the following four instruments: piano, cello, violin and flute. Considering that expressive performance synthesis for piano has been addressed in prior arts \cite{performancernn,hawthorne2018enabling}, we decide to experiment with the other three instruments here.
This reduces the total duration of the dataset to approximately 87 minutes, with 49, 30, 8 minutes for the cello, violin and flute, respectively. 

As the scale of the training set is small, we use the following simple data augmentation method to increase the data size---when cutting an audio clip into 5-second chunks, we largely increase the temporal overlaps between these chunks. Specifically, the overlaps between chunks are set to 4, 4.5, and 4.75 seconds for the cello, violin, and flute, respectively.
Although the augmented dataset would contain a large number of similar chunks, it still helps the model learn better, especially when we use a small hop size in the STFT to compute the spectrograms.

We remark that the training data we have for the flute is only composed of 3 audio clips that are 8 minutes long in total. It is interesting to test whether our model can learn to play the flute with such a small dataset.



\section{Experiment}

In training the model, we reserve 1/5 of the clips as the validation set for parameter tunning. 
We implement the model in Pytorch.
The model for the cello converges only after 8 hours of training, on  a singe NVIDIA 1080Ti GPU. 

We conduct two runs of subjective listening test to evaluate the performance
of the proposed model. 
156 adults participated in both runs. 41 of them music professionals or students majored in music. 
In the first run, we compare the result of our model with that of two MIDI synthesizers. \textbf{Synthesizer 1} uses an open-source one called Aria Maestosa, whereas  \textbf{Synthesizer 2}  uses Logic Pro, a commercial synthesizer.
The first synthesizer uses synthesized audio with digital signals, while the second one uses a library of real recordings.
Subjects are asked to listen to three 45-second audio pieces for each model (i.e., ours and the two synthesizers), without knowing what the model is. 
Each piece is the result for one instrument, for three pianorolls that are not in the training set. The subjects are recruited from the Internet and they are asked to rate the pieces in terms of the following metrics in a five-point Likert scale:
\begin{itemize}
\item Does the \textit{timbre} sound like real instruments?
\item \textit{Naturalness}: whether the performance is generated by human, not by machine?
\item How good is the \textit{audio quality}?
\item \textit{Emotion}: whether the performance expresses emotion?
\item The \textit{overall} perception of the piece.
\end{itemize} 
We average the ratings for the three pieces and consider that as the assessment for a subject for the result of each model.

\begin{figure}[t]
\begin{center}
\includegraphics[width=\columnwidth]{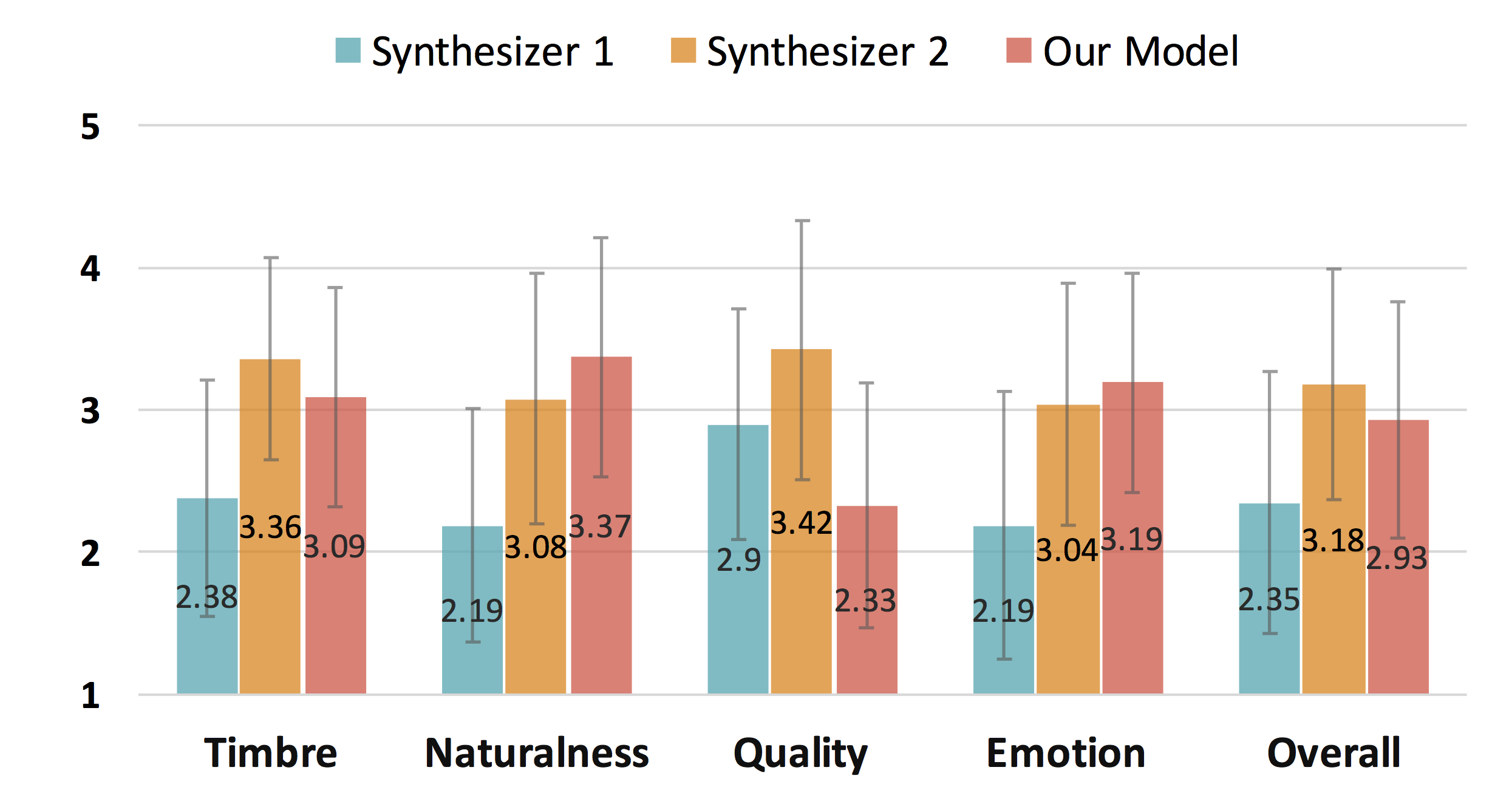}
\caption{The result of subjective evaluation for the first run, comparing the generated samples for three different instruments by our model and by two synthesizers.}
\label{fig:us3}
\end{center}
\end{figure} 

Figure \ref{fig:us3} shows the mean opinion scores (MOS). The following
observations are made. 
First, and perhaps the most importantly, the proposed model performs better than the two synthesizers in terms of \textit{naturalness} and \textit{emotion}. This shows that our model indeed learns the expressive attributes in music performance. 
In terms of \textit{naturalness}, the proposed model outperforms synthesizer 2 (i.e., Logic Pro) by 0.29 in MOS.
Second, in terms of \textit{timbre}, the MOS of the proposed model falls within those of the two synthesizers. This shows that the timbre of our model is better than soft synth, but it is inferior to that of real performance.
Third, there is still room to improve the sound quality of our model. Both synthesizers perform better, especially synthesizer 2. This also affects the ratings in \textit{overall}.
From the comments of the subjects, our model works much better in generating performance attributes like velocities and emotion, but the audio quality has to be further improved.

\begin{figure}[t]
\begin{center}
\includegraphics[width=\columnwidth]{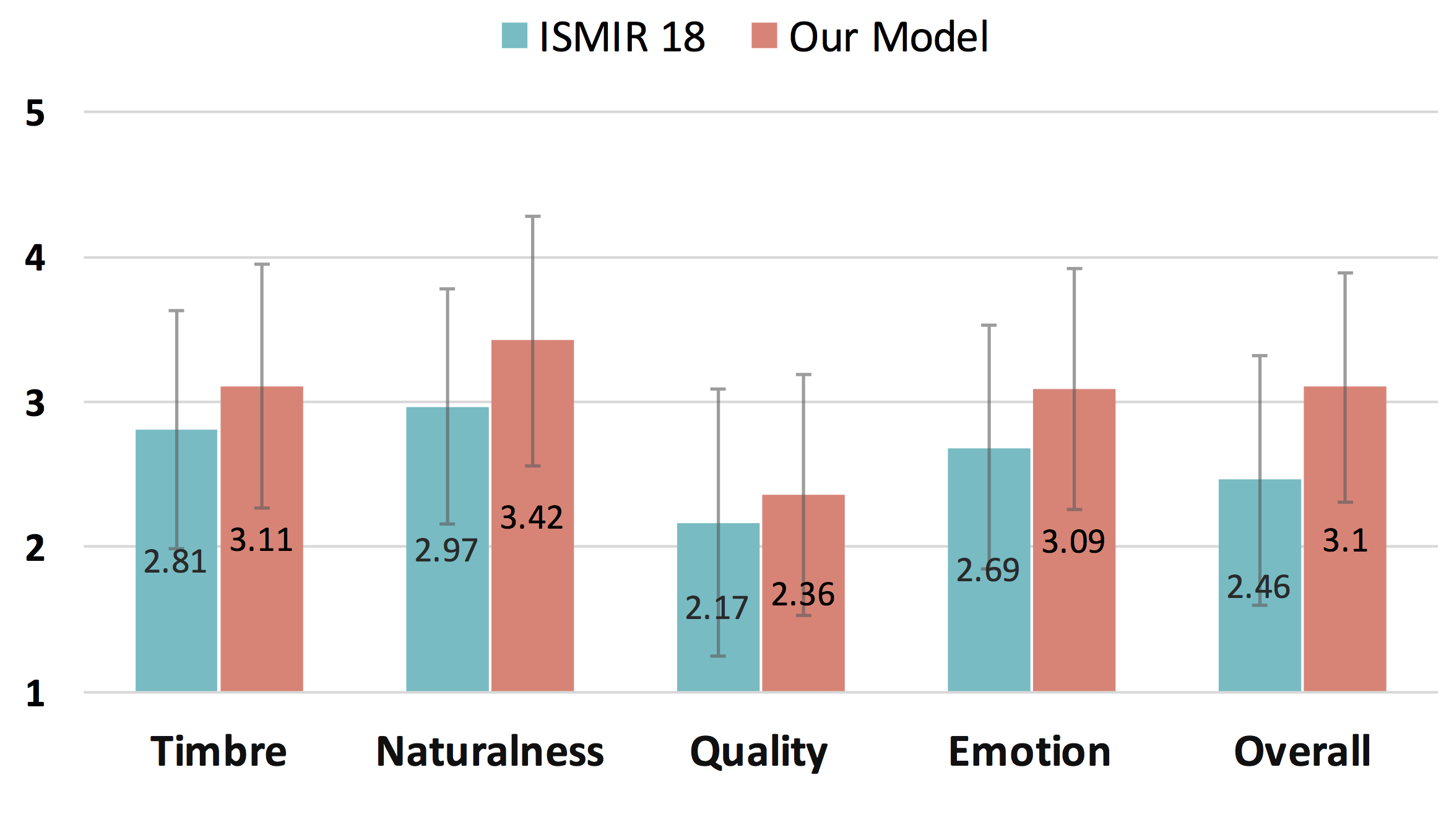}
\caption{The result of subjective evaluation for the second run, comparing the generated samples for the cello by our model and an existing model  \cite{manzelli18ismir}.}
\label{fig:us4}
\end{center}
\end{figure} 

In the second run, we additionally compare our model with the WaveNet-based model proposed by \cite{manzelli18ismir}. Because they only show the generated cello clips for two pianorolls on their demo website,
we only use our model to generate audio for the same two pianorolls for comparison. Figure \ref{fig:us4} shows that our model performs consistently better than this prior art in all the metrics. 
This result demonstrates how challenging it is for a  neural network to generate realistic audio performance, but it also shows that our model actually represents a big step forward.

\section{Conclusion and Discussion}

In this paper, we present a new deep convolutional model for score-to-audio music generation. The model takes as input a pianoroll and generates as output an audio clip playing that pianoroll. The model is trained with audio recordings of real performance, so it learns to render the pianorolls in an expressive way. Because the training is data-driven, the model can learn to play an instrument in different styles and flavors provided a training set of corresponding audio recordings. Moreover, as our model is fully convolutional, the training is fast and it does not require much training data. The user study shows our model achieves higher mean opinion score in naturalness and emotional expressivity than a WaveNet-based model and two off-the-shelf synthesizers, for generating solo performance of the cello, violin and flute.


We discuss a few directions of future research below. Firstly, 
the user study shows that our model does not perform well for timbre and audio quality. We may improve this by GAN training or a WaveNet-based decoder.

We can extend the \textit{U-Net+Encoder} architecture of the ContourNet to better condition the generation process. For example, we can use multiple encoders to incorporate different side information,
such as the second-by-second instrument activity \cite{hung18ismir} when the input has multiple instruments, and the intended playing technique for each note. In addition, the use of datasets with performance-level annotations (e.g., \cite{guitarset}) 
also makes it possible to condition and better control the generation process. 

To facilitate learning the score-to-audio mapping, it is important to have some ways to objectively evaluate the generated result. We plan to explore the following metrics: i) a pre-trained instrument classifier for the timbre aspect of the result, ii) a pre-trained pitch detector for the pitch aspect, and iii) psychoacoustic features \cite{psysound} such as spectral 
tonalness/noisiness for the perceptual audio quality.


The proposed multi-band residual (MBR) blocks can be used for other audio generation tasks such as audio-to-audio, text-to-audio or image/video-to-audio generation. Furthermore, the proposed `translation + refinement' architecture may find its application in translation tasks in other domains.
Compared to existing two-stage methods \cite{wang2018pix2pixHD,karras18iclr}, 
the proposed model may have some advantages since the training is done end-to-end. 

Because the MIDI/audio files of the MusicNet dataset 
are aligned, our model does not learn the \emph{timing} aspect of real performances. 
And, we do not generate multi-instrument music yet.
These are to be addressed in the future.

Finally, there are many important tasks 
to realize an ``AI performer,'' e.g. to model the personality, style, and intended emotion of a performer. 
We expect relevant research to flourish in the near future.

\bibliographystyle{aaai}
\bibliography{ref}

\end{document}